\documentclass[12pt]{article}
\begin{document}
\thispagestyle{empty}
%\vspace*{3cm}
\begin{center}
{\Large\sc Klein-Gordon Thermal Equation\\
 For A Preplanckian Universe}

\bigskip
{Miroslaw Kozlowski$^{\rm a}$}

{Janina Marciak-Kozlowska$^{\rm b,*}$}
\end{center}

\bigskip
\noindent{$^{\rm a}$~} Institute of Experimental Physics, Warsaw
University,
   Hoza~69, 00-681 Warsaw, Poland

\noindent{{$^{\rm b}$~} Institute of Electron Technology,
   Al.~Lotnik\'{o}w~32/46, 02-668 Warsaw, Poland

\hbox to 5cm{\hsize=5cm\vbox{\ \hrule}}\par

\noindent{{$^{\rm *}$~}Author to whom correspondence should be addressed.

\vspace{2cm}
\begin{abstract} In this paper the quantum hyperbolic equation formulated
in~\cite{1} is applied
to the study of the propagation of the initial thermal state of the
Universe. It is shown that the propagation depends on the barrier height.
The Planck wall potential is introduced, $V_P=\hbar/8t_P=1.25 \;10^{18}$
GeV where $t_P$ is a Planck time.
 For the barrier height $V<V_P$ the master thermal equation is {\it the  modified Klein-Gordon 
equation}, and for barrier height $V>V_P$ the master equation is
{\it the Klein-Gordon equation}. The solutions of both type equations for Cauchy
boundary conditions are discussed.

 \end{abstract}

\newpage
\section{Introduction}
In this paper the thermal behaviour of a Planck gas in the presence of
the potential barrier is investigated. The generalized quantum hyperbolic
heat transport equation formulated in~\cite{1} is applied to the
study of the propagation of the initial thermal state of the Universe. It
will be shown that the propagation depends on the barrier height.  The
thermal information on the Beginning is carried through the distorted
thermal waves. But the undistorted thermal information is completely
diminished for the time of the order of a Planck time.

In paper the possibility of the motion ``up the stream of time'' (in
spirit of W.~Thompson and J.~C.~Maxwell) is discussed~\cite{2}. It will
be shown that only hyperbolic heat transport equation guarantees the
possibility of this motion. This possibility does not exists with Fourier
type (parabolic) heat transport equation.

\section{Klein-Gordon equation for a Planck gas}

On time scales of Planck time, black holes of the Planck mass spontaneously
come into existence. Via the process of Hawking radiation, the black hole can
then evaporates back into energy. The characteristic time scale for this to
occur happens to be approximately equal to Planck time~$t_p$. Thus the Universe at
$10^{-43}$ seconds in age was filled with Planck gas i.e. gas of massive
particles all with masses equal Planck mass $M_p$. In the following we will
describe the thermal properties of the Planck gas in the field of the
potential~$V$.

As was shown in~\cite{1} the thermal properties of the Planck gas i.e. for $t<t_p$ can be
described by hyperbolic quantum heat transport equation~\cite{1}, viz:
\begin{equation}
t_p\frac{\partial^2 T}{\partial t^2}+\frac{M_p}{\hbar}\frac{\partial
T}{\partial t}+\frac{2VM_p}{\hbar^2}T=\nabla^2T.\label{eq1}
\end{equation}
In equation~(\ref{eq1}) $t_p$ denotes Planck time, $M_p$ is the Planck mass and
$V$ denotes the potential energy.

For the uniform Universe it is possible to study only one-dimensional heat
transport phenomena. In the following we will consider the thermal properties
of a Planck gas in constant potential $V=V_0$. In that case the one
dimensional quantum heat transport equation has the form:
\begin{equation}
\frac{1}{c^2}\frac{\partial^2 T}{\partial t^2}+
\frac{M_p}{\hbar}\frac{\partial T}{\partial t}+
\frac{2V_0M_p}{\hbar^2}T=\frac{\partial^2T}{\partial x^2},\label{eq2}
\end{equation}
where formula for $t_p=\hbar/M_pc^2$ was used~\cite{1}. In equation~(\ref{eq2})
$c$ - denotes the light velocity. As $c\neq\infty$ we can not omit the second
derivative term and consider only Fokker-Planck equation:
\begin{equation}
\frac{M_p}{\hbar}\frac{\partial T}{\partial
t}+\frac{2V_0M_p}{\hbar^2}T=\frac{\partial^2 T}{\partial x^2},\label{eq3}
\end{equation}
for heat diffusion in the potential energy $V_0$, or free heat diffusion:
\begin{equation}
\frac{\partial T}{\partial t}= \frac{\hbar}{M_p}\frac{\partial^2
T}{\partial x^2}.\label{eq4}
\end{equation}
It occurs that only if we retain the second derivative term we have the chance
to study the conditions in the Beginning.

Some implications of the forward and backward properties of the parabolic
heat diffusion equation were beautifully described by
J.~C.~Maxwell~\cite{2}.

As can be easily seen the second derivative term in equation~(\ref{eq1})
carriers the memory of the initial state which occurred at time $t=0$. If
we pass with $c\rightarrow\infty$ we lost the possibility to study the
influence of the initial conditions at the present epoch as it is
explained by J.~C.~Maxwell~\cite{2}. It means that by limiting procedure
$c\rightarrow\infty$ we cut off the memory of the Universe.

For hyperbolic quantum heat transport equation~(\ref{eq2}) we seek a solution
of the form:
\begin{equation}
T(x,t)=e^{-t/2t_p}u(x,t).~\label{eq5}
\end{equation}
After substitution of equation~(\ref{eq5}) to equation~(\ref{eq2}) one obtains:
\begin{equation}
\frac{1}{c^2}\frac{\partial^2 u}{\partial t^2}-\frac{\partial^2
u}{\partial x^2} +qu=0,~\label{eq6}
\end{equation}
where
\begin{equation}
q=\frac{2V_0M_p}{\hbar^2}-\left(\frac{M_pc}{2\hbar}\right)^2.\label{eq7}
\end{equation}
In the following we shall consider positive values of $V_0$, $V_0\geq0$,
i.e. we shall consider the potential barriers and steps.

The structure of the Eq.~(\ref{eq6}) depends on the sign of the parameter
$q$. Let us define the Planck wall potential, i.e. potential for which
$q=0$. From equation~(\ref{eq7}) one obtains:
\begin{equation}
V_P=\frac{\hbar}{8t_P}=1.25 \;10^{18} \;{\rm GeV},\label{eq8}
\end{equation}
where $t_P$ is a Planck time. For $q<0$, i.e.  when $V_0<V_P$ Eq.~(\ref{eq6})
is {\it the modified Klein-Gordon equation}~(MK-G)~\cite{3}. For the Cauchy
initial condition:
\begin{equation}
u(x,0)=f(x), \qquad \frac{\partial u(x,0)}{\partial t}=g(x),\label{eqn8}
\end{equation}
and the solution of Eq.~(\ref{eq5}) has the form~\cite{3}:
\begin{eqnarray}
u(x,t)&=&\frac{f(x-ct)+f(x+vt)}{2}\nonumber\\
&&\mbox{}+\frac{1}{2c}\int_{x-ct}^{x+ct}g(\zeta)I_0
\left[\sqrt{-q(c^2t^2-(x-\zeta)^2)}\right]d\zeta\label{eq9}\\
&&\mbox{}+{\frac{(c\sqrt{ -q}) t}
{2}}\int_{x-ct}^{x+ct}f(\zeta)\frac{I_1\left[
\sqrt{-q(c^2t^2-(x-\zeta)^2)}\right]}{\sqrt{c^2t^2- (x-\zeta)^2}}
d\zeta.\nonumber
\end{eqnarray}
In equation~(\ref{eq9}) $I_0$, $I_1$ denotes the Bessel modified function of
the zero and first order respectively.

When $q>0$, i.e for $V_0>V_P$ equation~(\ref{eq6}) reduces to {\it the
thermal Klein-Gordon Equation}~(K-GE).

For the Cauchy initial condition~(\ref{eqn8}) the solution of K-GE can be
written as~\cite{3}:
    \begin{eqnarray}
    u(x,t)&=&\frac{f(x-ct)+f(x+ct)}{2}\nonumber\\
    &&\mbox{}+\frac{1}{2c}\int_{x-ct}^{x+ct}g(\zeta)J_0
    \left[\sqrt{q(c^2t^2-(x-\zeta)^2)}\right]d\zeta\label{eq10}\\
    &&\mbox{}-\frac{(c\sqrt{ q }) t} {
    2}\int_{x-ct}^{x+ct}\frac{J_1
    \left[\sqrt{q(c^2t^2-(x-\zeta)^2)}\right]}
    {\sqrt{c^2t^2-(x-\zeta)^2}}d\zeta.\nonumber
    \end{eqnarray}
The case for $q=0$ was discussed in paper~\cite{1} and it describes the
distortionless quantum thermal waves. Both solutions~(\ref{eq9}) and
(\ref{eq10}) exhibit the domains of dependence and influence for the
modified Klein-Gordon equation and Klein-Gordon equation. These domains,
which characterize the maximum speed, $c$, at which the thermal
disturbance travels are determined by the principal terms of the given
equation (i.e. the second derivative terms) and do not depend on the
lower order terms. It can be concluded that these equations and the wave
equation have identical domains of dependence and influence. Both
solutions~(\ref{eq9}) and (\ref{eq10}) represents the distorted thermal
waves in the field of potential barrier or steps $V$.

\section{Conclusions}
In the paper the thermal behaviour of a Planck gas in the presence of a
potential barrier is investigated. It was argued that the hyperbolic
quantum heat transport equation offers the possibility for the study of
the thermal history of the Universe up to the Beginning, but the
information is transmitted through the distorted thermal waves. It was
shown that for a barrier height $V<V_P$ the quantum heat transport
equation is the {\it modified Klein-Gordon equation}. For a barrier
height $V>V_P$ the quantum heat transport equation is {\it the
Klein-Gordon equation}. It is quite interesting to observe that only for
$t_P\neq0$ the Planck wall has the finite height.

\end{document}